\documentclass[12pt]{article} 
\usepackage{epsf}
\usepackage{graphicx}
\usepackage{pazh}

\begin{document}
\baselineskip 21pt

\title{\bf Stellar populations and galaxy evolution in the NGC 80 group}

\author{\bf O.K. Sil'chenko$^1$ and V.L. Afanasiev$^2$}

\affil{
{\it  Sternberg Astronomical Institute of the Moscow State University, Moscow, Russia}$^1$\\ 
{\it Special Astrophysical Observatory of the Russian Academy of Sciences, Nizhnji
Arkhyz, Russia}$^2$}

\vspace{2mm}
\sloppypar 
\vspace{2mm}

\noindent
We have studied the central parts of seven early-type galaxies --
the members of the X-ray-bright galaxy group NGC 80 -- by means
of integral-field spectroscopy at the Russian 6m telescope. We
searched for signatures of synchronous evolution of the group
galaxies. The following results have been obtained. Five 
galaxies have revealed old stellar populations in the bulges,
with the SSP-equivalent ages from 10 to 15 Gyr. A moderate-luminous
S0 galaxy IC 1548 demonstrates consequences of recent star
formation burst: the SSP-equivalent age of the bulge is 3 Gyr,
that of the nucleus -- 1.5 Gyr. It is also in this galaxy that
we have found a circumnuclear polar gaseous disk which changes
smoothly to counterrotating one at radii larger than
$3^{\prime \prime}$ (1 kpc).
Probably, IC 1548 had suffered an interaction with external
gas accretion which might also provoke the central star formation
burst. In the giant E0 galaxy NGC 83 which is projected close to
the group center but has a line-of-sight velocity redshifted
by 600 km/s with respect to the group $v_r$, we have observed
a compact massive stellar-gaseous disk with the radius of some
2 kpc demonstrating current star formation. Consequently,
NGC 83, just as IC 1548, has the young stellar population
in the center. We speculate that a small subgroup leaded by
NGC 83 is in process of infalling into the old massive group
around NGC 80.

\clearpage

\section{INTRODUCTION}

\noindent
Evolution of galaxies is directed both by external and internal
factors. Internal factors include mainly dynamical instabilities
and depend on mass and angular momentum of a galaxy. External
factors are determined by the galaxy environments -- other
galaxies, intergalactic medium -- and can be either gravitational
or hydrodynamical mechanisms. Galactic groups are the best place
to study the external factors of galaxy evolution. Many of them
have X-ray halo of hot gas; this high-pressure intergalactic medium
can strip the own gas reservoirs of spiral galaxies and transform
them into lenticulars. Gravitational interactions are provided
by the tight neighborhood.  Galaxy velocity dispersions inside the 
groups are not so large, of order of the intrinsic stellar velocity
dispersions of early-type galaxies, and cannot prevent development
of tidal effects. Numerical simulations proved that tidal interactions
affect not only external parts of galaxies but also stimulate bar
formation in the centers which in turn re-distribute matter along
the radius. Secular evolution can even change the morphological
type of a galaxy by causing a growth of a bulge by resulting in
gas inflow and subsequent star formation in the center; a whole 
class of bulges, named `pseudobulges', may be formed rather
recently by such a mechanism (Kormendy and Kennicutt 2004).

If several galaxies are settled within similar environments in a center 
of a group and mutually interact gravitationally, it seems probable
that the re-shaping of their structures and formation of new central
stellar components should be synchronous. Observationally, it means
that the stellar ages would be the same, and the spatial star
distributions would be similar. If observations reveal synchronous
evolution of the central parts of the group galaxies, it would be
an argument in favour of dominance of external mechanisms; if no --
we should put attention to variance of internal conditions in the
galaxies under consideration. We have already studied 5 nearby groups
with the Multi-Pupil Fiber Spectrograph (MPFS) of the Russian
6m telescope; in each group we have observed central parts of
2--3 centrally located galaxies. In Leo I (Sil'chenko et al. 2003a)
and in the groups around NGC~5576 (Sil'chenko et al. 2002) and
NGC~3169 (Sil'chenko and Afanasiev 2006) we have found similar
properties of the stellar populations in the circumnuclear stellar
disks: they all have been formed rather recently, 1--3 Gyr ago;
and in 3 dominant galaxies of Leo I they are almost co-spatial.
In the Leo Triplet, on the contrary, the stellar ages and kinematics
in the centers of NGC~3623 and NGC~3627 were quite different 
(Afanasiev and Sil'chenko 2005), and we concluded that the galaxies
of the Triplet had met recently, not earlier than 1 Gyr ago.
The only group with the X-ray halo among those studied by us is
Leo II (Afanasiev and Sil'chenko 2007); its two central galaxies,
NGC~3607 and NGC~3608, have both old kinematically decoupled stellar
structures, but their mean stellar ages are different, 10 and 6 Gyr
respectively. In this paper we present results of our study of
stellar populations in galaxies of another massive group with 
X-ray hot gas -- the group of NGC~80.

A group of galaxies around the giant lenticular galaxy NGC 80 is
a rich and massive one. Ramella et al. (2002) in their catalog of
galactic groups mentioned 13 members within 2 magnitudes from the
brightest one, Dell'Antonio et al. (1994) estimated the number
of galaxies in the group as 21, and Mahdavi et al. (2004), with
more data, -- even as 45. The X-ray luminosity of the group, 
according to Mahdavi et al. (2000), is $\log L_x = 42.56 \pm 0.09$
($L_x$ is expressed in erg/s); it is rather high 
for a group. The center of the X-ray brightness distribution coincides
practically with NGC 80, the most luminous galaxy in the group.
According to Mahdavi et al. (2004), the velocity dispersion
of the galaxies in the NGC 80 group is 336 km/s and the systemic
velocity is $5771 \pm 48$ km/s; Dell'Antonio et al. (1994) rejected
some non-member galaxies and after that obtained 246 km/s and
$5663 \pm 51$ km/s, respectively. The central galaxy of the
group, the S0 NGC 80, has its own line-of-sight velocity of
5698 km/s which is close to the systemic velocity of the group;
it is consistent with NGC 80 being a dynamical center of the
group. Interestingly, another giant galaxy, a E0 NGC 83, having
the same luminosity as NGC 80 and being located very nearby
the group center, has a line-of-sight velocity exceeding the systemic
group velocity by more than 500 km/s. One more luminous galaxy
in the group is a spiral galaxy NGC 93. Bothun and Schommer (1983)
observed NGC 93 at the 21 cm, in the neutral-hidrogen emission line,
and they found that the galaxy is very massive, with the rotation
velocity of 317 km/s, and `anemic', that means very red and with the
low $M_{HI}/L_B$ ratio -- just as massive spiral galaxies of the
Virgo cluster. Spiral galaxies in clusters are thought to be `anemic'
because of effect of the surrounding hot intergalactic medium --
the most probable cause is the stripping of neutral hydrogen from
the outer disks of galaxies by ram pressure. Since the group of
NGC 80 is rich by hot gas, NGC 93 may be also stripped in such 
a manner.

We have made integral-field spectroscopic observations of the central
parts of the most luminous galaxies in the group, NGC 80, NGC 83, and
NGC 93, as well as of E-galaxy NGC 79 and S0-galaxies NGC 86, IC 1541,
and IC 1548; the stellar population properties and stellar and gaseous
(if the gas is present) kinematics have been studied. The galaxies
studied are distributed from the very center (NGC 80 and NGC 83) to
the most peripheric parts, at 0.5 Mpc from the group center (IC 1548
and IC 1541). Their global parameters taken from the NED and HYPERLEDA
are given in the Table 1.

{\normalsize
\begin{table*}
\caption[ ] {Global parameters of the galaxies studied}
% %\begin{center}
\begin{flushleft}
\begin{tabular}{lccccccc}
\hline\noalign{\smallskip}
% % &  \\
NGC(IC) & 80 & 83 & 79 & 93 & 86 & (1541) & (1548) \\
Morph. type (LEDA$^1$) & SA$0-$ & E & E & S? Sab
& Sa & S0 & S0 \\
$D_{25},\, ^{\prime}$  (LEDA) & 1.82 & 1.29 & 0.81 & 1.35 & 0.76 & 0.76
& 0.69 \\
$B_T^0$ (LEDA) & 12.98 & 13.22 & 14.66 & 13.66 & 14.35 & 15.15 & 15.20\\
$M_B$ (LEDA) & --21.6 & --21.6 & --19.8 & --20.8 & --20.2 &
--19.5 & --19.4 \\
$(B-V)_e$ (LEDA) & 1.07 & 1.12 & -- & 1.16 & -- & -- & --  \\
$(U-B)_e$ (LEDA) & 0.66 & 0.60 & -- & 0.61 & -- & -- & -- \\
$V_r$, km/s (NED$^2$) & 5698 & 6227 & 5485 & 5380 & 5591 & 5926 & 5746 \\
Distance, Mpc & \multicolumn{7}{c}{77 (Mahdavi \& Geller 2004)} \\
$PA_{phot}$ (LEDA) & $4.5^{\circ}$ & $112^{\circ}$ &
$163^{\circ}$ & $49^{\circ}$ & $8^{\circ}$ &
$36^{\circ}$ & $78^{\circ}$  \\
$\sigma _*$, km/s (LEDA) & 260 & 262 & 201 & -- & -- & -- & 154 \\
\hline
\multicolumn{8}{l}{$^1$\rule{0pt}{11pt}\footnotesize
NASA/IPAC Extragalactic Database}\\
\multicolumn{8}{l}{$^2$\rule{0pt}{11pt}\footnotesize
Lyon-Meudon Extragalactic Database}
\end{tabular}
\end{flushleft}
\end{table*}
}

\section{OBSERVATIONS AND DATA REDUCTION}

\noindent
The central parts of all galaxies inder consideration, 
$16^{\prime \prime} \times 16^{\prime \prime}$, have been
observed with the MPFS at the prime focus of the Russian 6m telescope
(for the spectrograph description, see Afanasiev et al. 2001).
Two spectral ranges have been explored, the green one 4150--5650~\AA\
and the red one 5850--7350~\AA, with the reciprocal dispersion
of 0.75~\AA\ and the spectral resolution of about 3~\AA. The
detector was CCD EEUV$2048 \times 2048$. The green spectra have
been used to derive stellar kinematics and to calculate Lick
indices H$\beta$, Mgb, and $\langle \mbox{Fe} \rangle$. In the red range we
measured baricenter positions of the emission lines H$\alpha$
and [NII]$\lambda$6583, prominent in the spectra of NGC 83,
to derive the ionized-gas kinematics. The details of our approach
can be looked for in, e.g., Sil'chenko (2006) and Sil'chenko and
Afanasiev (2006). The spectral observation log is given in the
Table 2.

\begin{table*}
\caption[ ] {Spectral observations with the MPFS}
% \begin{center}
\begin{flushleft}
\begin{tabular}{llccc}
\hline\noalign{\smallskip}
NGC & Date &  T(exp), min & Spectral range, \AA & $FWHM_*^{\prime \prime}$ \\
\hline\noalign{\smallskip}
NGC 80 & 30.09.2003 & 90  & 4150--5650 & 1.7 \\
NGC 83 & 8.10.2004 & 120 & 4150--5650 & 1.2 \\
NGC 83 & 30.09.2005 & 80 & 5800--7300 & 3 \\
NGC 93 & 19.09.2006 & 80 & 4150--5650 & 1.5 \\
NGC 79 & 19.09.2006 & 120 & 4150--5650 & 1.5 \\
NGC 86 & 17.08.2007 & 90 & 4150--5650 & 1.4 \\
IC 1541 & 18.08.2007 & 90 & 4150--5650 & 1.4 \\
IC 1548 & 18.08.2007 & 120 & 4150--5650 & 1.4 \\
\hline
\end{tabular}
% \end{center}
\end{flushleft}
\end{table*}

\section{STELLAR POPULATIONS IN THE CENTERS OF THE GALAXIES}

\noindent
We studied already the stellar population in the center of NGC 80
with the previous version of the MPFS (Sil'chenko et al. 2003b). We
reported the discovery of a chemically distinct nucleus and a
ring of intermediate-aged stars with the radius of 
$5^{\prime \prime}-7^{\prime \prime}$; both these `secondary'
structures are embedded into the old bulge, with the mean stellar
age of 10 Gyr. The new data presented here have higher spectral
resolution and larger field of view; however the new radial index
profiles are in agreement with the older ones (Fig. 1). This fact
confirms our estimate of the Lick index measurement accuracy
and demonstrate stability of our calibration into the standard
Lick index system: as it is known, the spectral resolution of the
Lick data is about 8~\AA, and we observed with the spectral 
resolution of 5~\AA\ before 1998 and with that of 3~\AA\ now.

\begin{table}
\caption[ ]{Lick indices and the properties of the nuclear stellar populations}
% \begin{center}
\begin{flushleft}
\begin{tabular}{l|c|c|c|c|c|c|}
\hline\noalign{\smallskip}
NGC & H$\beta$ &  Mgb & $\langle \mbox{Fe} \rangle $ 
& T, Gyr & [Z/H] & [Mg/Fe] \\
\hline\noalign{\smallskip}
NGC 80 & 1.66 & 5.00 & 2.94 & 6 & $+0.5$ & $+0.29$  \\
NGC 83 & 1.87 & 5.20 & 2.94 & 4 & $+0.6$ & $+0.34$ \\
NGC 93 & 1.90 & 4.36 & 3.08 & 4 & $+0.4$ & $+0.12$ \\
NGC 79 & 1.15 & 5.22 & 3.16 & 15 & $+0.3$ & $+0.22$ \\
NGC 86 & 1.77 & 4.10 & 2.78 & 8 & $+0.2$ & $+0.17$ \\
IC 1541 & 1.48 & 4.76 & 2.94 & 12 & $+0.2$ & $+0.22$ \\
IC 1548 & 3.28 & 3.40 & 2.68 & 1.5 & $+0.7$ & $+0.06$ \\
\hline
\end{tabular}
% \end{center}
\end{flushleft}
\end{table}

\begin{table}
\caption[ ] {Lick indices and the stellar population properties in the bulges}
% \begin{center}
\begin{flushleft}
\begin{tabular}{l|c|c|c|c|c|c|}
\hline\noalign{\smallskip}
NGC & H$\beta$ &  Mgb & $\langle \mbox{Fe} \rangle$ 
& T, Gyr & [Z/H] & [Mg/Fe] \\
\hline\noalign{\smallskip}
NGC 80 & 1.76 & 4.50 & 2.81 & 7--8 & $+0.3$ & $+0.23$  \\
NGC 83 & 2.02 & 5.69 & 2.84 & 3 & $+0.7$ & $+0.47$ \\
NGC 93 & 1.63 & 3.98 & 2.62 & 10 & $0.0$ & $+0.20$ \\
NGC 79 & 1.52 & 5.04 & 2.97 & 12 & $+0.3$ & $+0.25$ \\
NGC 86 & 1.61 & 3.72 & 2.95 & 12 & $+0.1$ & $+0.01$ \\
IC 1541 & 1.20 & 3.79 & 2.32 & 15 & $0.0$ & $+0.28$ \\
IC 1548 & 2.42 & 2.81 & 2.18 & 3 & $-0.1$ & $+0.17$ \\
\hline
\end{tabular}
% \end{center}
\end{flushleft}
\end{table}

Tables 3 and 4 contain the Lick indices and parameters of the
stellar populations in all the galaxies studied here which have
been obtained by confronting our Lick indices to the models
of simple stellar populations (SSP) by Thomas et al. (2003).
They are presented separately for the unresolved nuclei and 
for the rings between the radii of $4^{\prime \prime}$ and 
$6^{\prime \prime}$ which we consider to be the bulge regions.
At the redshift of the NGC 80 group the linear distances from
the center which we have used to probe the bulge properties
correspond to about 2 kpc. The accuracy of the Lick indices
both for the nuclei and for the bulges is about 0.15~\AA,
the corresponding accuracies of the abundances and ages are
0.1 dex for [m/H], 0.05 dex for [Mg/Fe], and 1 Gyr for young
ages ($T<8$ Gyr) and 2 Gyr for old ages. The Lick index
H$\beta$ given in the Tables 3 and 4 is corrected for the
emission contribution in the galaxies where the emission
presence is evident: in NGC 83, NGC 93, and IC 1548. For
NGC 83 we have calculated the corrections through the
equivalent width of the emission line H$\alpha$ measured
in the red spectra of the galaxy: we have used the empiric
relation by Stasinska and Sodre (2001) for the composite
gas excitation mechanism, 
$EW(\mbox{H}\beta emis)=0.25EW(\mbox{H}\alpha emis)$.
For NGC 93 and IC 1548 we used the mean relation between
the emission lines H$\beta$ and [OIII]$\lambda$5007 obtained
for the centers of early-type galaxies by Trager et al. (2000),
$EW(\mbox{H}\beta)=0.6EW(\mbox{[OIII]}\lambda 5007)$. The
whole procedure of deriving the stellar population parameters
was as follows. We have taken the SSP models by Thomas et al. (2003)
because they are calculated for four values of the Mg/Fe ratio.
To derive SSP-equivalent ages and metallicities of the stellar 
populations, we firstly compared our data and the models 
at the diagram H$\beta$ versus [MgFe]
([MgFe]$\equiv \sqrt{\mbox{Mgb} \langle \mbox{Fe} \rangle}$):
this diagram allows, on one hand, to separate age and metallicity
effects and to solve the age-metallicity degeneracy, and on
the other hand, it is rather insensitive to the Mg/Fe ratio
which is unknown in advance. After determining the age of the
stellar population, we confront Mgb and 
$\langle \mbox{Fe} \rangle \equiv$(Fe5270+Fe5335)/2 to derive
the abundance ratio [Mg/Fe]: in the models by Thomas et al.
(2003) the range of [Mg/Fe] from -0.3 to +0.5 is covered.
According to modern models of chemical evolution, the
magnesium-to-iron ratio is a good indicator of the star
formation duration: the solar [Mg/Fe] indicates continuous
star formation during several Gyr, [Mg/Fe]$=+0.3 - +0.5$
corresponds to a brief star formation burst, stopping
before the bulk of intermediate-mass SNeIa begin to explode.

Figures 2 and 3 present the diagnostic diagrams for the nuclei
and the bulges of the `second-rank' members of the group: the
elliptical NGC 79, the early-type spirals NGC 93 and NGC 86,
and the edge-on lenticulars IC 1541 and IC 1548. Four of them
have rather old stellar populations in the centers: 
in the nuclei the ages range from 5 to 15 Gyr, the bulges 
are homogeneously old, $T$ is from 10 to 15 Gyr. Only
one lenticular, IC~1548, is outstanding: it demonstrates
a rather strong emission line [OIII]$\lambda$5007 both
in the nucleus and in the bulge, and the SSP-equivalent
ages are 1.5 Gyr for the nucleus and 3 Gyr for the bulge.
The most easy explanation is that IC 1548 is only recently
accreted onto the group and has transformed from a spiral
galaxy by hot intragroup medium pressure. However in our
sample IC 1548 is not the S0 galaxy farthest from the group
center; IC 1541 is the farthest one, and it has no emission
lines in the spectra, and no signs of recent star formation.
Perhaps, more probable reason of the recent star formation 
burst in IC 1548 is tidal interaction with the neighboring 
galaxy, PGC 1662109, that may be accompanied by gas accretion,
because PGC 1662109 is a blue gas-rich barred spiral galaxy
and is only 50 kpc from IC 1548. Since the bulge of IC 1548
demonstrates the higher age and Mg/Fe ratio then the nucleus,
we can suggest that the star formation burst starting over 
the extended central region some 3-4 Gyr ago, has persisted
rather long time only in the very nucleus where the stellar
metallicity has grown to the very high value for such 
low-luminosity galaxy, [Z/H]$=+0.7$. Interestingly, the
nuclei of NGC 93 and IC 1541 being rather old are also
chemically distinct; it means that these galaxies had
nuclear secondary star formation bursts, but long ago.
We conclude that the main part of the NGC 80 group 
gathered together at $z > 0.5$ and is a rather old
structure.

Now let us discuss two central galaxies of the group.
They have the same luminosity, the same colour, and
look very similar: both are large and round. However,
NGC 80 is a lenticular, and NGC 83 is an elliptical.
Fig. 4 shows diagnostic Lick-index diagrams for these 
galaxies. The bulge age for NGC 80 mentioned in the
Table 4 is in fact a formal mean value including the
old, $T\sim 10$ Gyr, bulge and a rather young, 
$T\sim 5$ Gyr, stellar ring. As we have already noted 
earlier (Sil'chenko et al. 2003b), the origin of the
rather young stellar ring with the radius of 2.5 kpc
in the center of NGC 80 is a puzzle: usually ring 
stellar structures are related to bar resonances, and
this galaxy is perfectly axisymmetric, it has no neither 
bar nor oval distortion. But the center of NGC 83, the
giant elliptical galaxy, is even more impressive.
Both the nucleus and the circumnuclear area within 
$8^{\prime \prime}$ from the center demonstrate the
SSP-equivalent ages not more than 3--4 Gyr,
and within $5^{\prime \prime}$ from the center the
intense emission line H$\alpha$ indicates current star
formation.

\section{GASEOUS DISKS IN NGC 83 and IC 1548}

Table 5 gives kinematical results, both for stars and for
ionized gas in all 7 galaxies observed. The stellar velocity
dispersion are calculated separately for the nuclei and
for the bulges -- just as the Lick indices in the previous
section. When the orientations of the kinematical major axes
change along the radius due to a possible non-axisymmetry effect,
the Table 5 gives the range of values measured between
$R=2^{\prime \prime}$ and $R=5^{\prime \prime}$. The angular
rotation velocities in the Table 5 refer to the innermost
regions with rigid-body rotation, $R<2^{\prime \prime}$.

\begin{table*}
\caption[ ] {Kinematical parameters}
% \begin{center}
\begin{flushleft}
\begin{tabular}{l|c|c|c|c|c|}
\hline\noalign{\smallskip}
NGC & $V_{\mbox{sys}}, km/s $ & $\sigma _*$ (nuc) & $\sigma _*$ (bulge) 
& $PA_{kin}$ & $\omega _0 \sin i$, km/s/arcsec \\
\hline\noalign{\smallskip}
NGC 80 (stars) & 5746 & 192 & 140 & $30^{\circ} - 45^{\circ}$ & $6 \pm 8$ \\
NGC 83 (stars) & 6218 & 170 & 113 & $114^{\circ}$ & 62 \\
NGC 83 (gas) & -- & -- & -- & $116^{\circ} \pm 4^{\circ}$ & 90 \\
NGC 93 (stars) & 5539 & 166 & 124 & $50^{\circ} - 60^{\circ}$ & 57 \\
NGC 79 (stars)& 5408 & 156 & 120 & $\sim 270^{\circ}$ & $6 \pm 6$ \\
NGC 86 (stars) & 5611 & 130 & 92 & $11^{\circ} - -3^{\circ}$ & 47 \\
IC 1541 (stars) & 5961 & 155 & 86 & $223^{\circ} - 213^{\circ}$ & 27 \\
IC 1548 (stars) & 5661 & 140 & 110 & $76^{\circ} - 56^{\circ}$ & 37 \\
IC 1548 (gas) & -- & -- & -- & $349^{\circ} - 276^{\circ}$ & 43 \\
\hline
\end{tabular}
% \end{center}
\end{flushleft}
\end{table*}

In two early-type galaxies we have found central gaseous
disks and have estimated their orientation and rotation
parameters.

A giant elliptical galaxy NGC~83 has peculiar NIR radiation 
excess detected by the IRAS and a large amount of molecular
gas (Wiklind et al. 1995, Young 2005). Young (2005) noted
a two-horn profile of the integrated CO emission line, with
the peak separation of 417 km/s, and concluded that the
molecular gas was confined to a compact massive disk, but
could not definitely judge about the possible current star
formation in this disk. We have observed NGC~83 with the
MPFS in the red spectral range and have found strong emission
lines H$\alpha$ and [NII]$\lambda$6583 within $5^{\prime \prime}$
from the center; the intensity ratio of H$\alpha$ to
[NII] rises along the radius though not achieving 2 which
is a characteristic value when the gas is excited by young
stars (so called `HII-type' excitation). It looks like the
nucleus being a LINER while the outer parts of the gaseous
disk demonstrate a combination of the LINER-type and
HII-type gas excitation.

Figure 6 presents 2D line-of-sight (LOS) velocity fields
for the stars and ionized gas in the center of NGC~83
according to the MPFS data. They reveal regular rotation
both of the stars and of the ionized gas, with the rather
high projected velocity of about 200 km/s. This value agrees
with the width of the CO-line profile (Young 2005). The
isophotes look round, without any signs of diskiness; it means
that the inclination of the embedded stellar disk is lower
than, say, $40^{\circ}$. The visible ellipticilty of the
red (dust) ring in the center of NGC~83 (Young 2005) implies
the inclination of $26^{\circ} - 37^{\circ}$. If it is so
indeed, the deprojected rotation velocity of the circumnuclear
disk in NGC~83 exceeds 350 km/s.

The gas velocity field of the S0-galaxy IC~1548 (Fig. 7) 
calculated by measuring the baricenter positions of the
[OIII]$\lambda$5007 emission line is even more interesting.
At the edge of the gaseous disk, at $R=3^{\prime \prime} -
4^{\prime \prime}$, the ionized gas counterrotates with
respect to the stars ($PA_0 (gas)\approx 260^{\circ}$
{\bf vs} $PA_0 (stars)\approx 70^{\circ}$), but in the
very center, within $R<2^{\prime \prime}$, the ionized gas
passes to polar orbits. We found such combinations of the
outer counterrotating gas and the inner polar gas before 
(Sil'chenko 2005, Sil'chenko \&\ Moiseev 2006); they were
also theoretically predicted for the gas inside a bar 
potential by Friedli \&\ Benz (1993).

\section{DISCUSSION}

\noindent
We have studied 7 early-type (E--Sab) galaxies which are
members of the massive X-ray galaxy group NGC 80, by means
of 2D spectroscopy at the 6m telescope. We have searched for
consequences of their synchronous secular evolution. Five
of seven galaxies have old bulges, with the mean stellar ages
of 10--15 Gyr. Concerning the central galaxy of the group,
NGC~80, we have found earlier (Sil'chenko et al. 2003b) that
it has circumnuclear structures of intermediate age, namely,
the nucleus and the ring with the radius of about 2 kpc, but
they are not younger than 5 Gyr. The `anemic' giant spiral
galaxy NGC~93, which neutral gas has been probably stripped 
away partly by ram pressure of the hot intragroup medium at 
the moment of galaxy infall, has the mean stellar age of the 
chemically distinct nucleus of 4 Gyr. If to relate its
nuclear star formation burst having produced the chemically
distinct nucleus with the galaxy infall into the group,
we would obtain that the moment of the main group gathering
occured about 4--5 Gyr ago.

However there are some early-type galaxies in this group
that possess rather young stellar populations. The S0-galaxy
IC~1548 reveals clear signatures of the recent central star
formation burst: the mean stellar age in the bulge is 3 Gyr,
and that in the nucleus is 1.5 Gyr. In the same galaxy we have
found a circumnuclear polar gaseous ring which warps smoothly
into the counterrotating gaseous disk coplanar with the stellar 
one in more outer parts. Probably, IC~1548 suffered strong
tidal interaction with the nearby late-type spiral galaxy and
perhaps even accreted gas from this neighbor; once acquiring the
gas, IC 1548 had to be provoked to the central star formation
by the same tidal interaction.

The giant E0 galaxy NGC~83 possesses a compact massive 
stellar-gaseous disk with the radius of some 2 kpc, very
fast rotating, with the noticeable current star formation.
Perhaps, all these properties are consequences of merging
a spiral or an irregular gas-rich galaxy, so called `minor
merger'. It is interesting that the line-of-sight systemic
velocity of NGC~83, a giant galaxy in the center of the
group, exceeds far the systemic LOS velocity of the group,
by more than two group velocity dispersion values. Only
a few nearby group members -- NGC 81, NGC 85, PGC 1327 --
have LOS velocities close to that of NGC~83. The majority
of the group members have LOS velocities that differ from
that of NGC~80 by no more than 200--300 km/s. We may suggest
that the poor subgroup of NGC~83 has been accreted by the
group of NGC~80 only recently. Was NGC~83 an elliptical
galaxy before the accretion event? Did its central gas
belong to NGC~83 before the subgroup accretion, or now
it is a result of merging? Several variants of answers
are possible.

\medskip
We thank Dr. A. V. Moiseev of SAO RAS for supporting
the observations at the 6m telescope.
The 6m telescope is operated under the financial support of
Science Ministry of Russia (registration number 01-43).
During the data analysis we have
used the Lyon-Meudon Extragalactic Database (LEDA) supplied by the
LEDA team at the CRAL-Observatoire de Lyon (France) and the
NASA/IPAC Extragalactic Database (NED) which is operated by the
Jet Propulsion Laboratory, California Institute of Technology,
under contract with the National Aeronautics and Space Administration.
Our study  is supported by the grant of the Russian Foundation for 
Basic Researches no. 07-02-00229a

\clearpage

\begin{figure}
\includegraphics[width=\hsize]{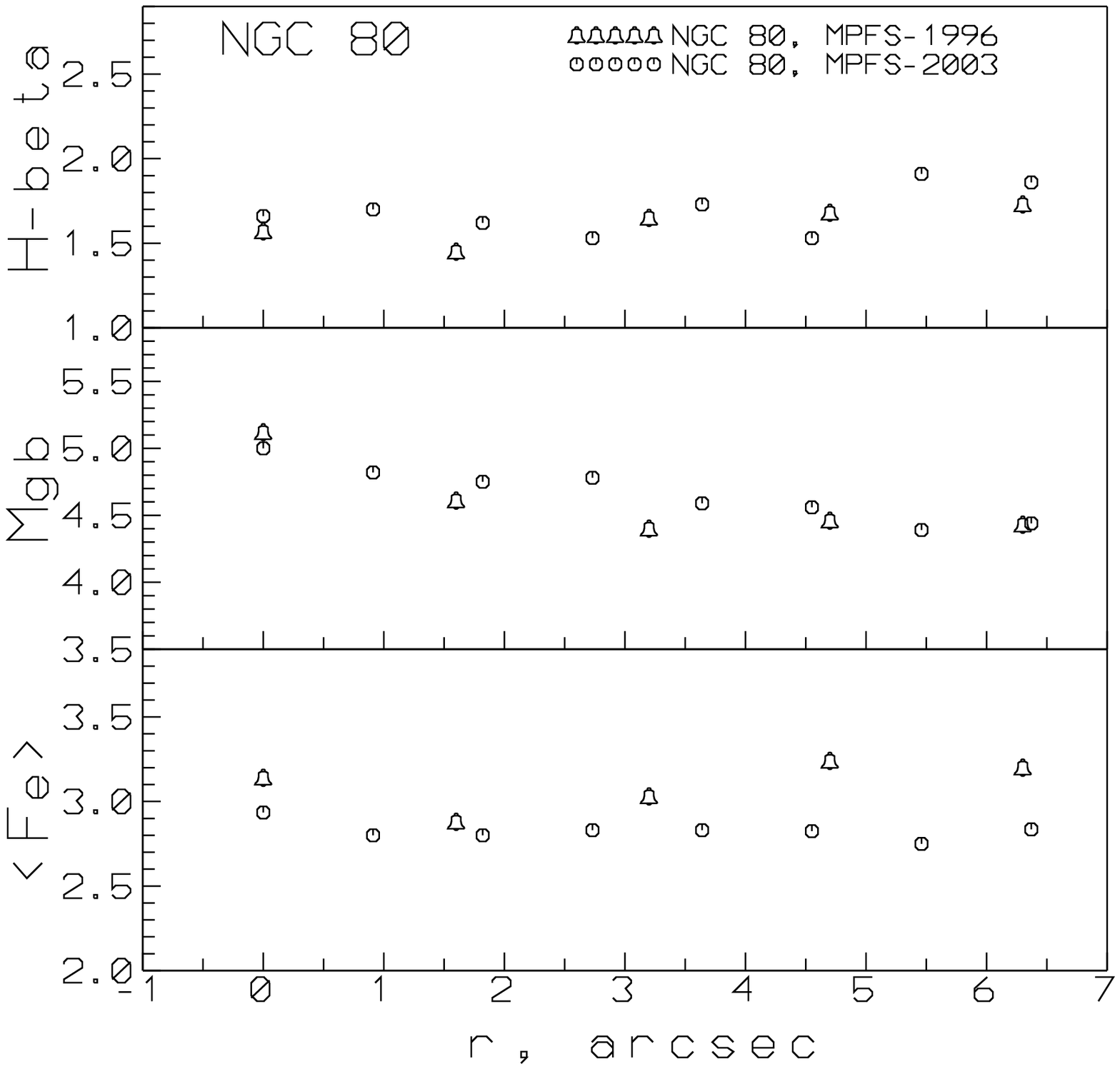}
\caption{
Radial variations of the azimuthally-averaged Lick indices
H$\beta$, Mgb, and $<\mbox{Fe}> \equiv$(Fe5270+Fe5335)/2
in NGC~80 according to our MPFS data of 1996 (bells) and 2003(circles).}
\end{figure}

\begin{figure}
\begin{tabular}{cc}
\includegraphics[width=0.45\hsize]{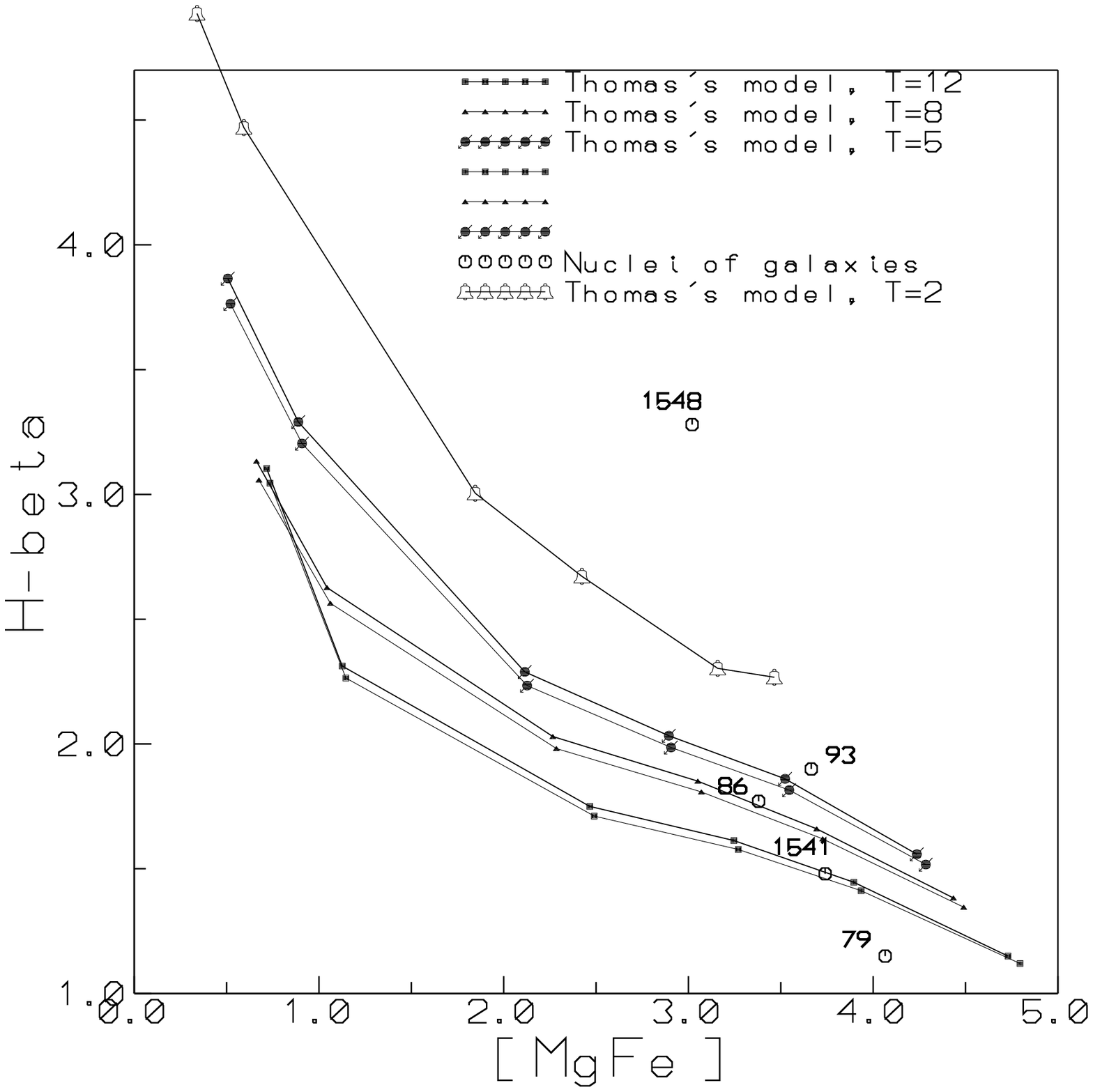} &
\includegraphics[width=0.45\hsize]{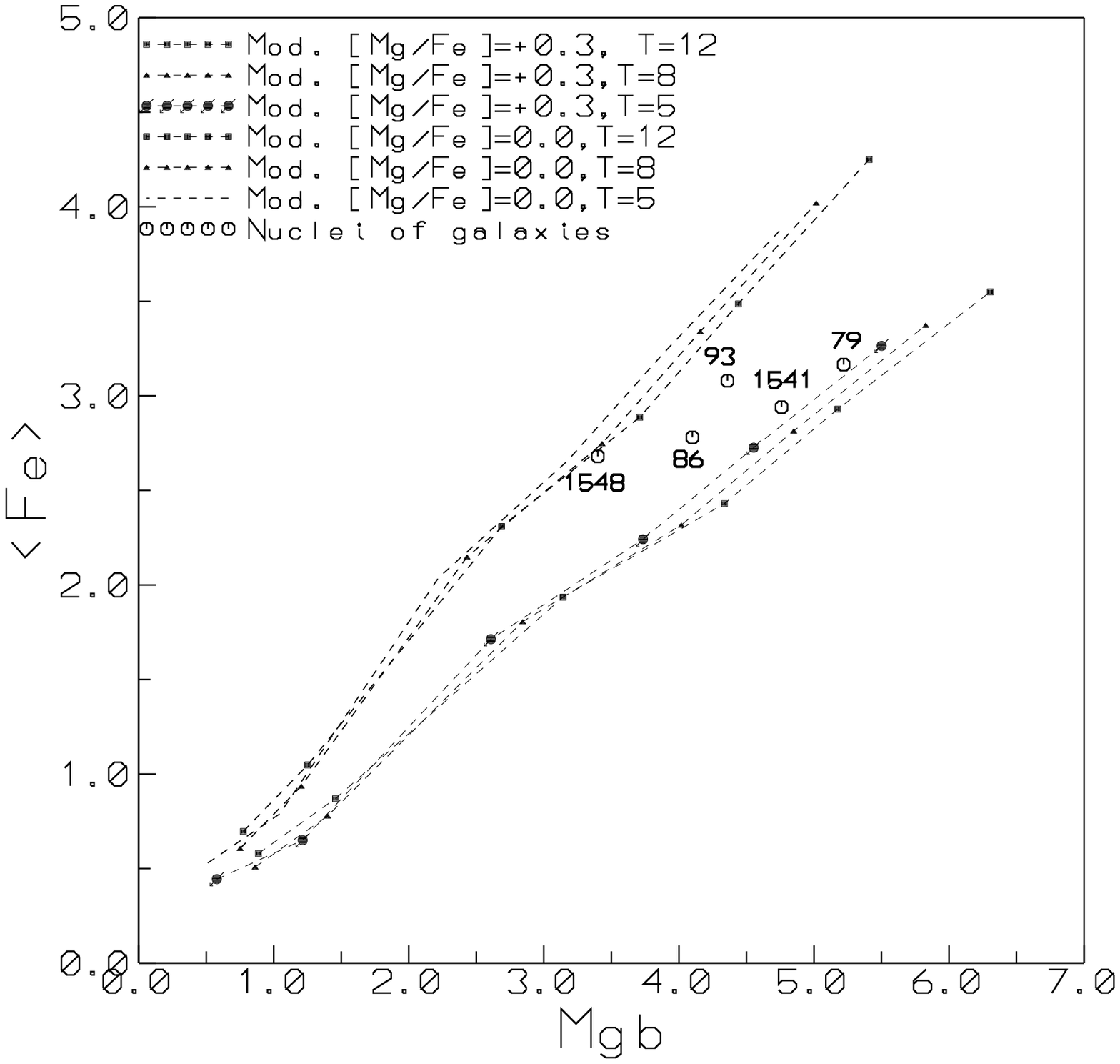}
\end{tabular}
\caption{
Diagnostic `index--index' diagrams for the nuclei of five off-center
group member galaxies. Model sequences of constant ages from Thomas
et al. (2003) are also plotted; the metallicities of the models  
are +0.67, +0.35, 0.00, --0.33, --1.35, --2.25, if one takes the small 
signs along the sequences from the right to the left.}
\end{figure}

\begin{figure}
\begin{tabular}{cc}
\includegraphics[width=0.45\hsize]{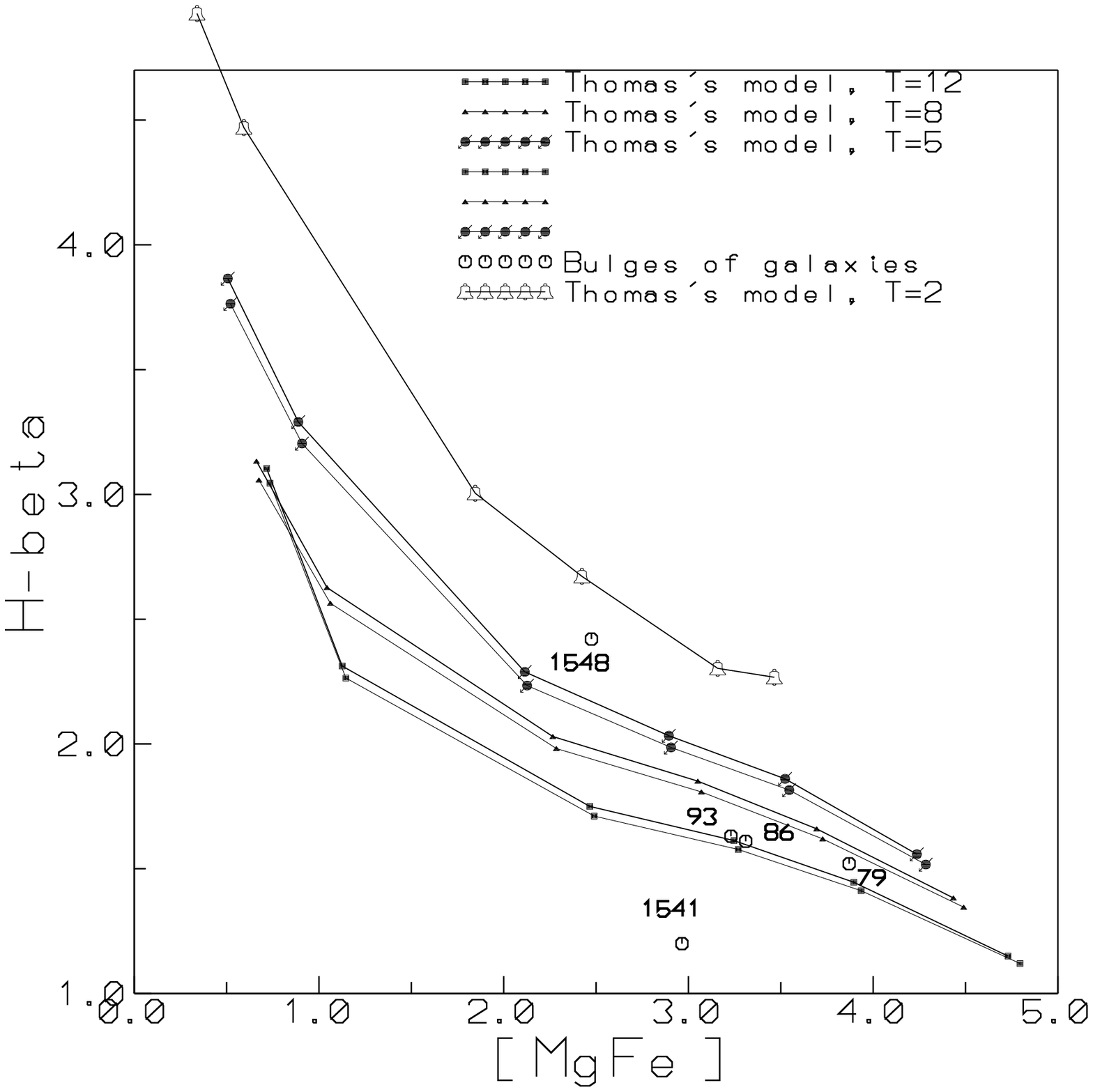} &
\includegraphics[width=0.45\hsize]{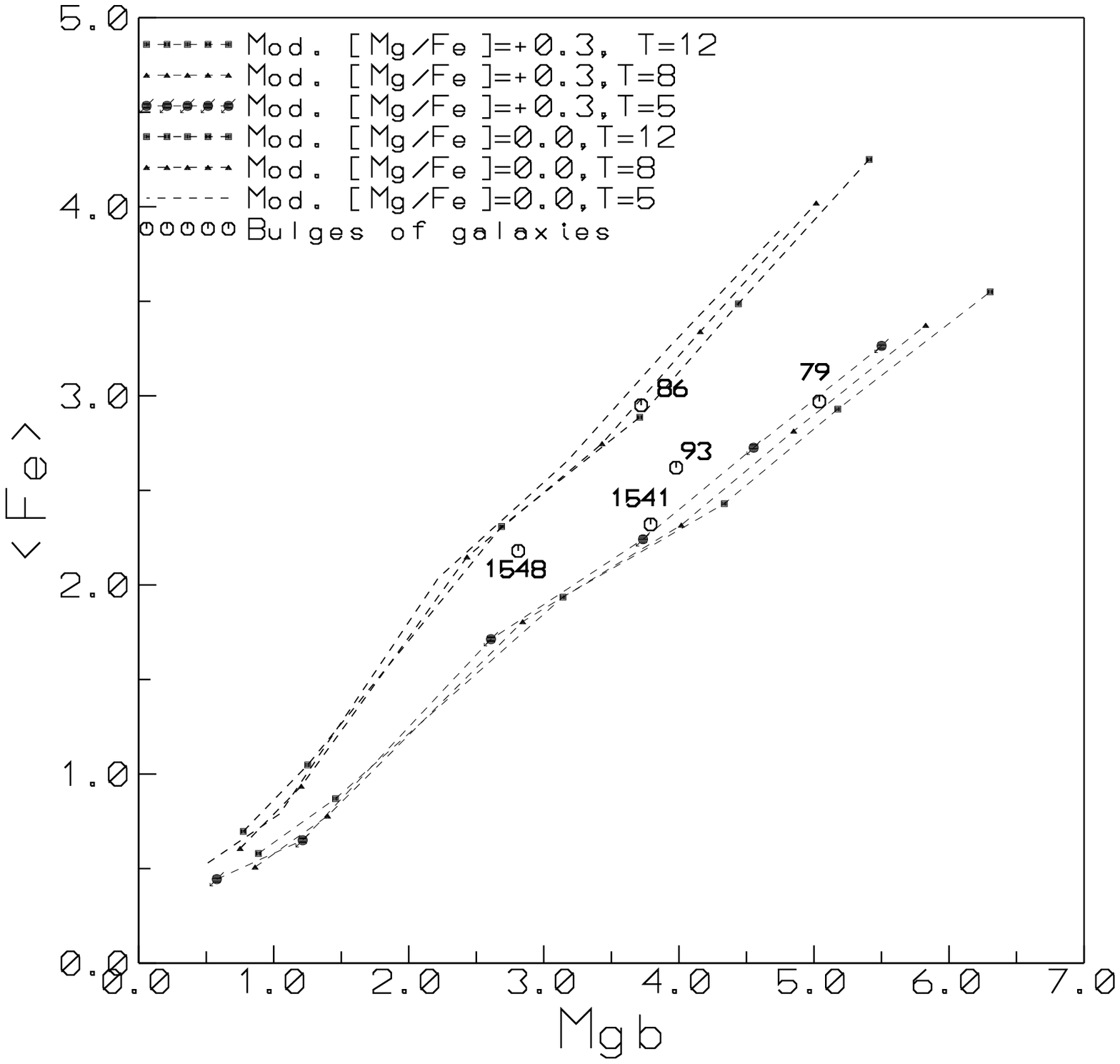}
\end{tabular}
\caption{
Diagnostic `index--index' diagrams for the bulges of five off-center
group member galaxies. Model sequences of constant ages from Thomas
et al. (2003) are also plotted; the metallicities of the models  
are +0.67, +0.35, 0.00, --0.33, --1.35, --2.25, if one takes the small 
signs along the sequences from the right to the left.}
\end{figure}

\begin{figure}
\begin{tabular}{cc}
\includegraphics[width=0.45\hsize]{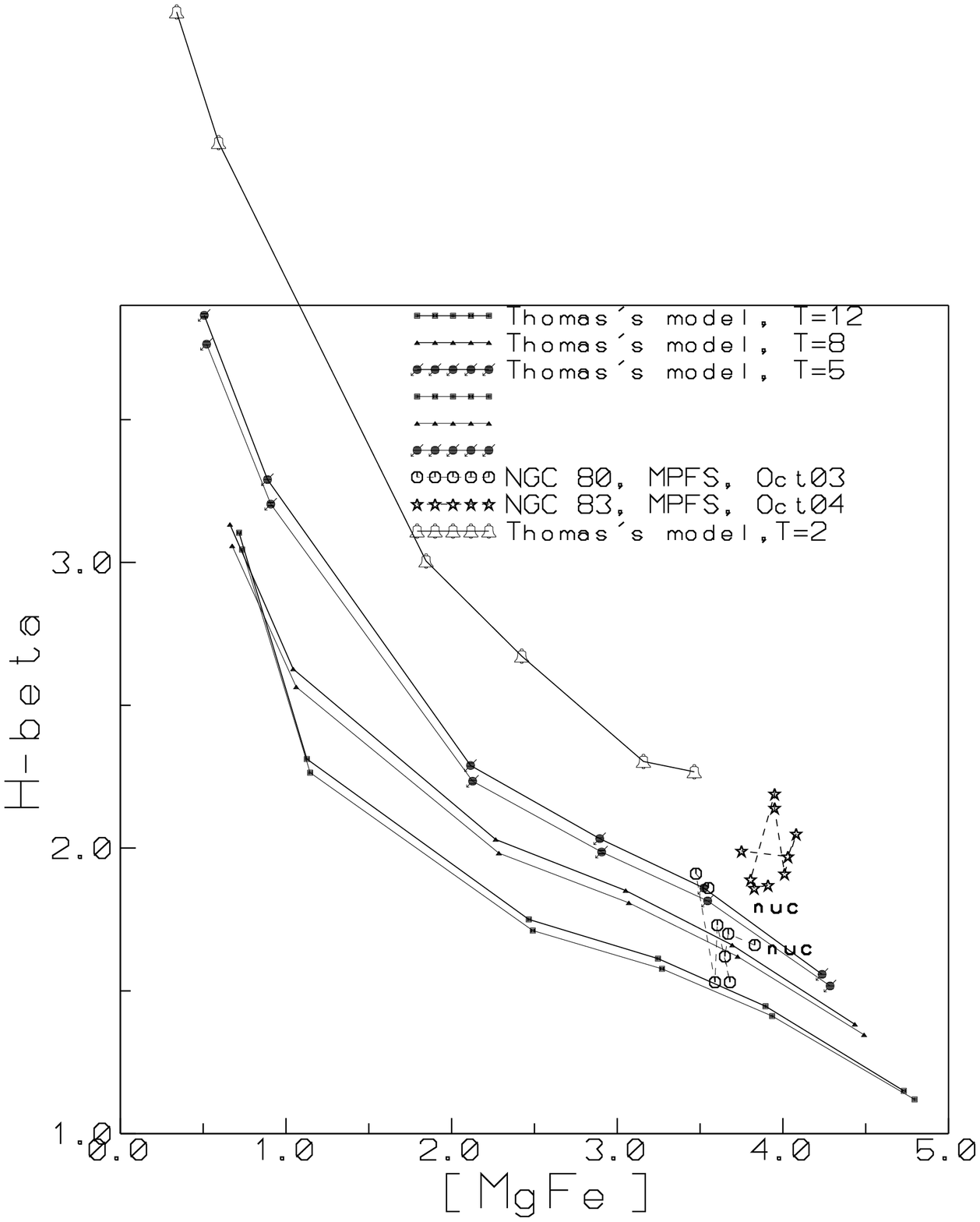} &
\includegraphics[width=0.45\hsize]{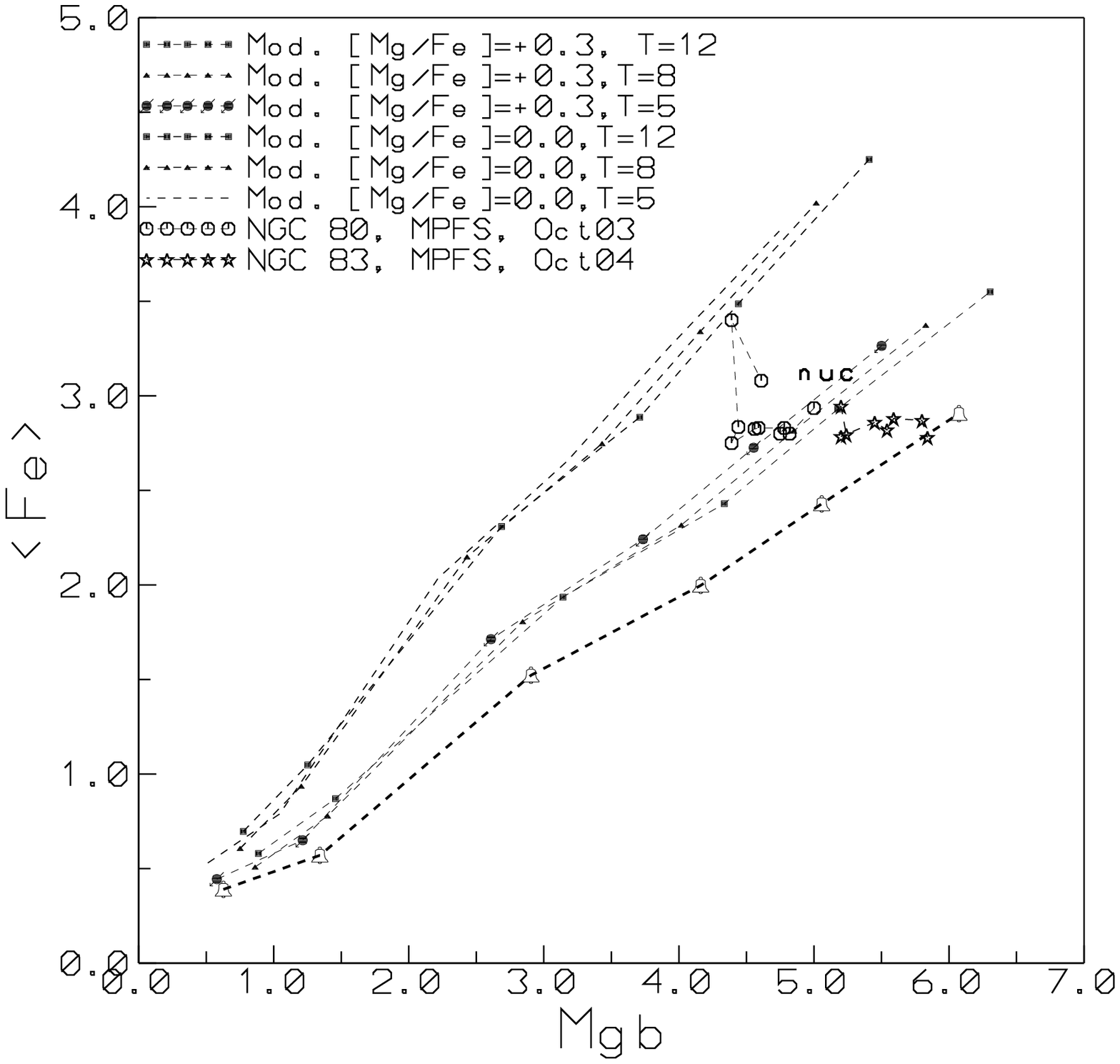}
\end{tabular}
\caption{
Diagnostic `index--index' diagrams for the azimuthally-averaged
index measurements in NGC 80 (circles) and NGC 83 (asteriks).
The measurements are made with the step of $1^{\prime \prime}$ along
the radii, are connected by the dashed lines; the nuclei are marked
by {\bf nuc}. The model sequences of constant ages from Thomas
et al. (2003) are also plotted; the metallicities of the models  
are +0.67, +0.35, 0.00, --0.33, --1.35, --2.25, if one takes the small 
signs along the sequences from the right to the left.}
\end{figure}

\begin{figure}
\includegraphics[width=\hsize]{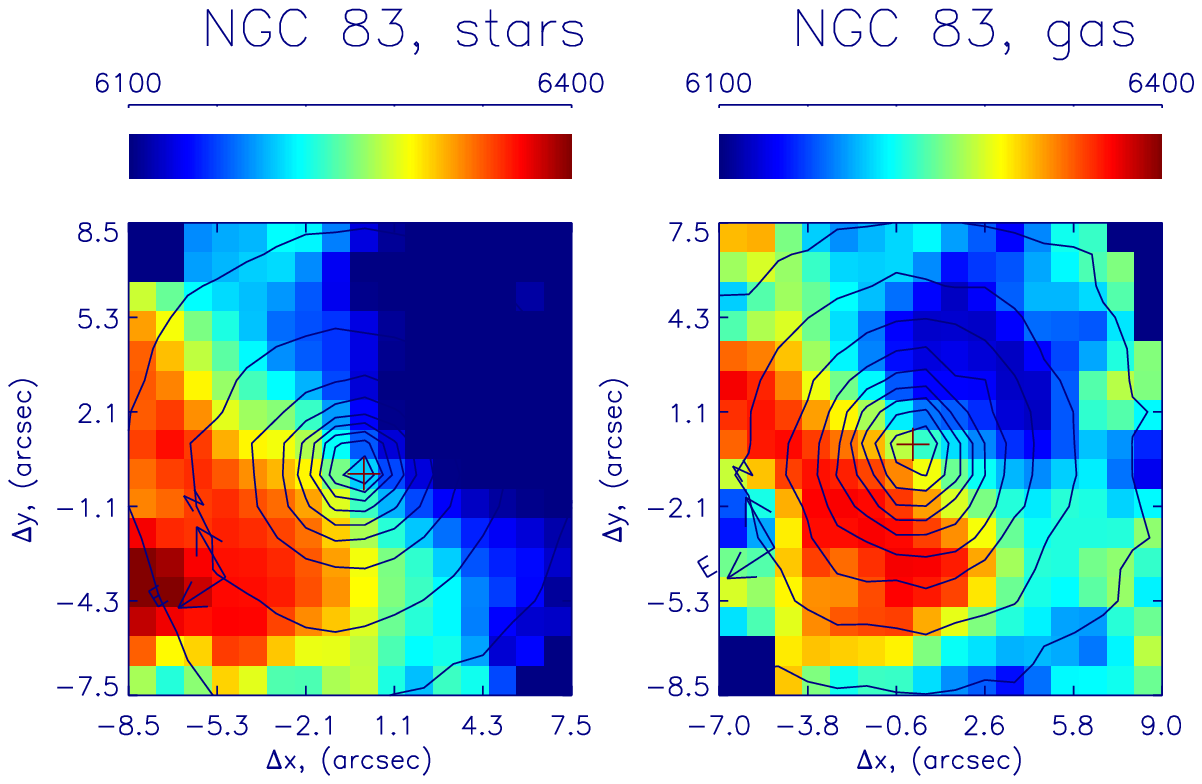}
\caption{Two-dimensional line-of-sight velocity fields in the center of
NGC 83: {\it left} -- the stellar component, {\it right} -- the
ionized gas, the isophotes superimposed.}
\end{figure}

\begin{figure}
\includegraphics[width=\hsize]{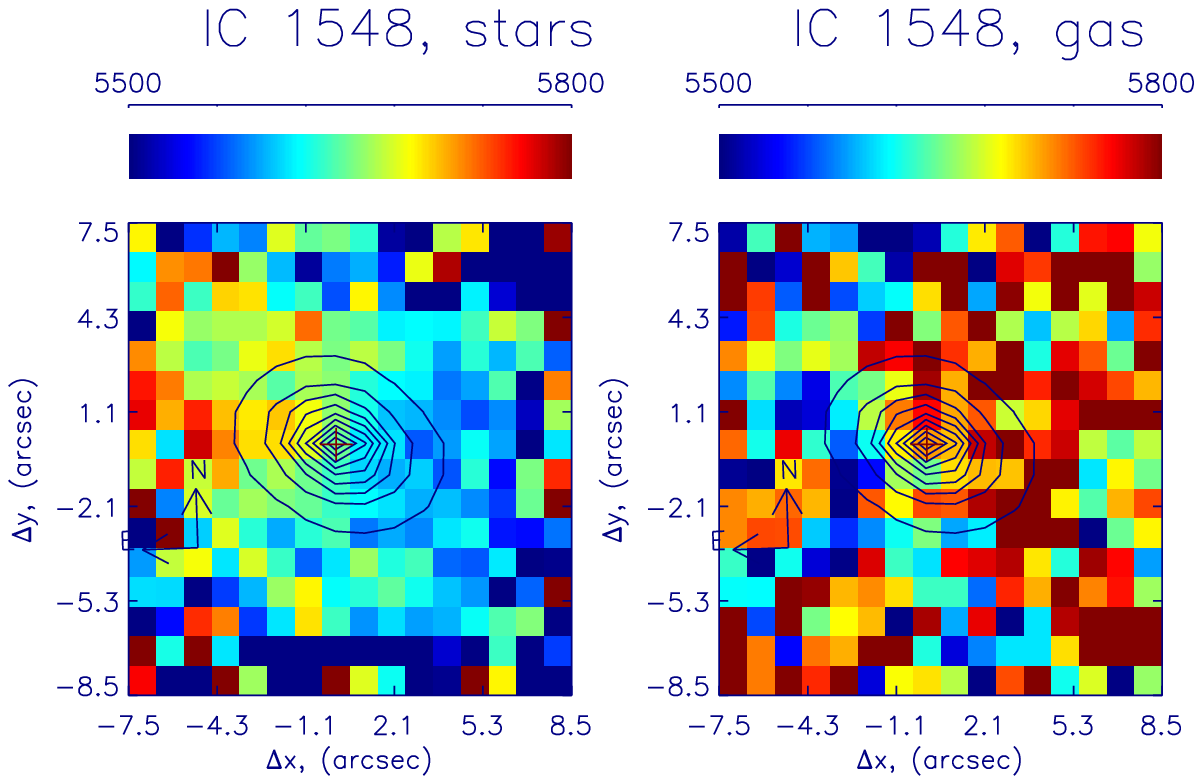}
\caption{Two-dimensional line-of-sight velocity fields in the center of
IC 1548: {\it left} -- the stellar component, {\it right} -- the
ionized gas, the isophotes superimposed.}
\end{figure}


\clearpage

\begin{references}

\reference{\it 
Afanasiev V.L., Dodonov S.N., Moiseev A.V.}, Proc. of the Conf. 
   "Stellar dynamics: from classic to modern", St. Petersburg, 2001/ 
    Eds. Osipkov L.P. and Nikiforov I.I., Saint Petersburg Univ. press, p.103

\reference{\it 
V.L. Afanasiev, O.K. Sil'chenko}, Astron. and Astrophys. {\bf 429}, 825 (2005)

\reference{\it 
 V.L. Afanasiev, O.K. Sil'chenko}, Astron. and Astrophys. Trans. {\bf 26}, 311 (2007)

\reference{\it 
G.D. Bothun, R.A. Schommer}, Astrophys. J. {\bf 267}, L15 (1983)

\reference{\it 
 I.P. Dell'Antonio, M.J. Geller, D.G. Fabricant}, Astron. J. {\bf 107}, 427 (1994)

\reference{\it 
D. Friedli, W. Benz}, Astron. Astrophys. {\bf 268}, 65 (1993)

\reference{\it 
J. Kormendy}, Astrophys. J. {\bf 257}, 75 (1982)

\reference{\it 
J. Kormendy, R.C. Kennicutt Jr.}, Ann. Rev. Astron. Astrophys. {\bf 42},
        603 (2004)

\reference{\it 
A. Mahdavi, M.J. Geller}, Astrophys. J. {\bf 607}, 202 (2004)

\reference{\it 
A. Mahdavi, H. Bohringer}, M.J. Geller, M. Ramella, Astrophys. J. 
   {\bf 534}, 114 (2000)

\reference{\it 
 M. Ramella, M.J. Geller, A. Pisano, L.N. da Costa}, Astron. J. {\bf 123},
 2976 (2002)

\reference{\it 
O.K. Sil'chenko}, Astron. Letters, {\bf 31}, 227 (2005)

\reference{\it 
O.K. Sil'chenko}, Astrophys. J., {\bf 641}, 229 (2006)

\reference{\it 
O.K. Sil'chenko, A.V. Moiseev}, Astron. J. {\bf 131}, 1336 (2006)

\reference{\it 
 O.K. Sil'chenko, V.L. Afanasiev}, Astron. Letters, {\bf 32}, 534 (2006)

\reference{\it 
O.K. Sil'chenko, V.L. Afanasiev, V.H. Chavushyan, J.R. Valdes}, 
      Astrophys.J. {\bf  577}, 668 (2002)

\reference{\it 
O.K. Sil'chenko, A.V. Moiseev, V.L. Afanasiev, V.H. Chavushyan, J.R. Valdes},
       Astrophys.J. {\bf  591}, 185 (2003a)

\reference{\it 
O.K.Sil'chenko, S.E. Koposov,, V.V. Vlasyuk, O.I. Spiridonova}, Astron. Rep., 
{\bf 47}, 88 (2003b)

\reference{\it 
G. Stasinska, I. Sodre Jr.}, Astron. Astrophys., {\bf 374}, 919 (2001)

\reference{\it 
 D. Thomas, C. Maraston, R. Bender}, MNRAS {\bf 339}, 897 (2003)

\reference{\it 
S.C. Trager, S.M. Faber, G. Worthey, J.J. Gonz\`alez}, Astron. J.,
{\bf 119}, 1645 (2000)

\reference{\it 
T. Wiklind, F. Combes, C. Henkel}, Astron. Astrophys. {\bf 297}, 643 (1995)

\reference{\it 
G. Worthey, S.M. Faber, J.J. Gonz\`alez, D. Burstein}, 
    Astrophys. J. Suppl. Ser. {\bf 94}, 687 (1994)

\reference{\it 
G. Worthey}, 1994, Astrophys. J. Suppl. Ser. {\bf 95}, 107 (1994)

\reference{\it 
L.M. Young}, Astrophys. J. {\bf 634}, 258 (2005)

\end{references}
\end{document}